%% file: main.tex
\documentclass[5p,times,twocolumn]{elsarticle}
\usepackage{physics}
\pdfoutput=1
\usepackage{tabularx}
\usepackage{siunitx}
\usepackage{xcolor}

\usepackage{hyperref}
\hypersetup{hidelinks}
\usepackage{csquotes}
\usepackage{amssymb}
\usepackage{pifont}
\newcommand{\cmark}{\ding{51}}%
\newcommand{\xmark}{\ding{55}}%
\usepackage{acro}
\input{sections/00_acro}

\graphicspath{{figures/}}
\usepackage{lipsum}
\usepackage{booktabs}
\usepackage{float}
\usepackage{comment}
\usepackage{subcaption}
\usepackage{framed} 
\usepackage{multicol} 
\usepackage{nomencl} 
\usepackage{xurl}
\usepackage{todonotes}

\usepackage{makecell}
\usepackage{afterpage}

\newenvironment{AllowDisplayBreaks}{\allowdisplaybreaks}{\ignorespacesafterend}
\newcommand\stxt[1]{\scriptscriptstyle\textup{#1}}

\usepackage{xcolor}
\usepackage{hyperref}

\definecolor{myColor}{RGB}{0, 0.0, 0} 

\hypersetup{
    colorlinks=true,
    linkcolor=myColor,
    citecolor=myColor,
    urlcolor=myColor,
}

\journal{Applied Energy}

\makeatletter
\def\ps@pprintTitle{
  \let\@oddhead\@empty
  \let\@evenhead\@empty
  \def\@oddfoot{}%
  \let\@evenfoot\@oddfoot
}
\makeatother

\begin{document}

\begin{frontmatter}

\title{Imitation learning with artificial neural networks for demand response with a heuristic control approach for heat pumps}

\author[1]{Thomas Dengiz}
\author[1]{Max Kleinebrahm}

\address[1]{Karlsruhe Institute of Technology (KIT), Institute for Industrial Production (IIP), Karlsruhe, Germany}

\begin{abstract}
\input{sections/00_abstract}
\end{abstract}


\end{frontmatter}


\input{sections/00_nomecl}
\input{sections/01_intro}

\input{sections/02_related_work}
\input{sections/03_optimization_problem_for_the_building}

\input{sections/04_immitation_learning_for_building_control}

\input{sections/05_results}

\input{sections/06_conclusion}

\input{sections/ending}

\thispagestyle{empty} 

 \bibliographystyle{elsarticle-num} 
 \bibliography{refs}





\end{document}

%% file: sections/00_acro.tex
\DeclareAcronym{PSC}{
    short = PSC,
    long  = Price Storage Control,
}

\DeclareAcronym{ML}{
    short = ML,
    long  = Machine Learning,
}

\DeclareAcronym{ANN}{
    short = ANN,
    long  = Artificial Neural Network,
}

\DeclareAcronym{MILP}{
    short = MILP,
    long  = Mixed Integer Linear Programming,
}

\DeclareAcronym{RL}{
    short = RL,
    long  = Reinforcement Learning,
}

%% file: sections/00_abstract.tex
\acresetall
The flexibility of electrical heating devices can help address the issues arising from the growing presence of unpredictable renewable energy sources in the energy system. In particular, heat pumps offer an effective solution by employing smart control methods that adjust the heat pump's power output in reaction to demand response signals. This paper combines imitation learning based on an artificial neural network with an intelligent control approach for heat pumps. We train the model using the output data of an optimization problem to determine the optimal operation schedule of a heat pump. The objective is to minimize the electricity cost with a time-variable electricity tariff while keeping the building temperature within acceptable boundaries. We evaluate our developed novel method, PSC-ANN, on various multi-family buildings with differing insulation levels that utilize an underfloor heating system as thermal storage. The results show that PSC-ANN outperforms a positively evaluated intelligent control approach from the literature and a conventional control approach. Further, our experiments reveal that a trained imitation learning model for a specific building is also applicable to other similar buildings without the need to train it again with new data. Our developed approach also reduces the execution time compared to optimally solving the corresponding optimization problem. PSC-ANN can be integrated into multiple buildings, enabling them to better utilize renewable energy sources by adjusting their electricity consumption in response to volatile external signals.

%% file: sections/00_nomecl.tex
\setlength{\nomitemsep}{-\parskip} 
\makenomenclature
\renewcommand*\nompreamble{\begin{multicols}{2}}
\renewcommand*\nompostamble{\end{multicols}}

\ExplSyntaxOn
\NewExpandableDocumentCommand{\strcase}{mm}
 {
  \str_case:nn { #1 } { #2 }
 }
\ExplSyntaxOff

\renewcommand\nomgroup[1]{%
  \item[\bfseries
    \strcase{#1}{
      {P}{Parameters}
      {A}{Acronyms}
      {V}{Variables}
      {I} {Indices}
    }%
  ]%
}

\begin{table*}[htb]
  \begin{framed}
    \nomenclature[A]{\acs{PSC}}{\acl{PSC}}
    \nomenclature[A]{\acs{ANN}}{\acl{ANN}}
    \nomenclature[A]{\acs{RL}}{\acl{RL}}
    \nomenclature[A]{\acs{ML}}{\acl{ML}}
    \nomenclature[A]{\acs{MILP}}{\acl{MILP}}

%
    \nomenclature[V]{$C$}{total cost}
    \nomenclature[V]{$x_t$}{modulation degree of the heat pump}

    \nomenclature[V]{$p_t$}{price of electricity}

    \nomenclature[V]{$T^{\stxt{Building}}_t$}{building temperature at time step $t$}
    \nomenclature[V]{$T^{\stxt{Outside}}_t$}{building temperature at time step $t$}
    \nomenclature[V]{$Q_t^{\stxt{HP}}$}{heat pump heat output}

    \nomenclature[V]{$h^{\stxt{switchedOff}}_t$}{binary variable indicating if the heat pump is switched off at time $t$}

    \nomenclature[V]{$h^{\stxt{on}}_{t}$}{binary variable indicating if the heat pump is running at time $t$}

    \nomenclature[V]{$\widetilde{x}_{t,pred}$}{predicted output of the ML algorithm for the heat pump at time $t$}

%
    \nomenclature[P]{$P^{\stxt{HPmax}}$}{maximum heating power of the heat pump}
    \nomenclature[P]{$\Delta t$}{time step duration}
    \nomenclature[P]{$T^{\stxt{min}}$}{minimum allowable building temperature}
    \nomenclature[P]{$T^{\stxt{max}}$}{maximum allowable building temperature}
    \nomenclature[P]{$Q_t^{\stxt{DemandSH}}$}{space heating demand at time step $t$}
    \nomenclature[P]{$Q_t^{\stxt{LossesSH}}$}{heat losses in space heating}
    \nomenclature[P]{$P_t^{\stxt{DemandEl}}$}{electrical power of the inflexible devices}
    \nomenclature[P]{$V^{\stxt{UFH}}$}{volume of underfloor heating system}
    \nomenclature[P]{$\rho^{\stxt{Concrete}}$}{density of concrete}
    \nomenclature[P]{$c^{\stxt{Concrete}}$}{specific heat capacity of concrete}
    \nomenclature[P]{$COP_t$}{coefficient of Performance of the heat pump}
    \nomenclature[P]{$mod^{\stxt{min}}$}{minimum modulation degree of the heat pump}
    \nomenclature[P]{$n^{\stxt{switchedOff}}$}{maximum number of starts for the heat pump}
    \nomenclature[P]{$k^{\stxt{switchedOff}}_t$}{number of starts for the heat pump until time t}


    \nomenclature[I]{$t$}{time slot}
    \nomenclature[I]{$Z$}{total number of time slots}

    \printnomenclature
  \end{framed}
\end{table*}

%% file: sections/01_intro.tex
\section{Introduction}
\label{sec_introduction}
The increasing share of volatile renewable energy sources like photovoltaic and wind energy in many countries makes flexible electrical loads vital to cope with their intermittent nature. Especially electrical heating devices like heat pumps (HP) coupled to thermal storage can provide flexibility to the energy system by using demand response \cite{PATTEEUW2015306}. To this end, novel intelligent control strategies are necessary that regulate the heat pump's electricity consumption based on external signals, like a time-variable electricity tariff. 

A promising approach to improve current control approaches is to include methods from the field of machine learning in the control procedures. Supervised learning approaches are mainly used for forecasting demand and generation load profiles for the control problem \cite{ANTONOPOULOS2020109899}. Thus, supervised learning is not used to control the devices' actions directly. Many studies from the literature use Reinforcement Learning (RL) to train an agent in a building environment to control flexible devices optimally. However, RL has some crucial drawbacks that limit its suitability for real-world applications in the energy sector. The main disadvantage is that RL needs a lot of time to train the model and thus requires a lot of interactions with the environment \cite{Dinh2022_2, Cao2020}. Further, as the power and energy systems have high safety requirements, online training of the agent may negatively affect the operation of the device if no domain knowledge is included in the control. 

Therefore, we introduce a novel control approach that combines imitation learning with domain knowledge of the control problem. Imitation learning is a paradigm for supervised learning in which a model learns 
by observing and mimicking the behavior of an expert \cite{Dinh2022_2}. The goal is to enable the trained model to perform a task or make decisions similar to those of the expert, even in situations it has not encountered before. 

Optimally reacting to the volatile energy generation requires the repeated solving of an optimization problem. Finding the optimal solution is equivalent to depicting a high-dimensional mapping between the inputs of the problem (demand, external signal, outside temperature, etc.) and the optimal control actions. As the same problem is solved for the same building recurrently with different inputs, a trained machine learning model can learn the mapping between inputs and outputs and thus imitate the exact solver. The trained model can then be amortized over multiple problem instances. Therefore, it can lead to a reduction in execution time. Imitation learning has shown promising results in the energy field \cite{Dinh2022_2, Pan2021}, and we investigate its applicability to intelligently controlling a heat pump for demand response. 

The remainder of the paper is organized as follows: In Sec.~\ref{sec_related_work_and_contribution}, we summarize the relevant literature and highlight the contribution of this paper. We define the optimization problem in Sec.~\ref{sec:Optimization problem of the building}. Our novel control approach is explained in Sec.~\ref{sec:Imitation learning for building control}. We show the results of our experiments to evaluate the introduced control approach in Sec.~\ref{sec:Results}. This paper ends with a conclusion and outlook in Sec.~\ref{sec:Conclusion}.

%% file: sections/02_related_work.tex
\section{Related work and contribution}
\label{sec_related_work_and_contribution}

\subsection{Related work}
\label{subsec_Related work}

\begin{table*} 
    \centering
    \caption{Comparison of relevant papers from the literature}
    \begin{tabular}{|l|c|c|c|c|c|}
        \hline
        & \makecell{Forecast-free \\ control} & \makecell{Learning from \\ optimal control} & \makecell{Embedded in smart \\ control heuristic} & \makecell{Compared to optimal \\ and smart control} & \makecell{Training data from \\ other buildings} \\
        \hline
        Javed et al., 2017 \cite{Javed2017}  & \xmark & \xmark & (\cmark) & \xmark & \xmark\\
        \hline
         Kim et al., 2020 \cite{Kim2020}  & \xmark & \cmark & \xmark & \cmark & \xmark\\
         \hline
         Dey et al., 2023 \cite{DEY2023}  & \cmark & \xmark & \xmark & (\cmark) & \xmark\\
         \hline
         Zou et al., 2020 \cite{Zou2020}  & \cmark & \xmark & (\cmark) & \xmark & \xmark\\

         \hline
         Dinh et al., 2022 \cite{Dinh2022_2}  & \cmark & \cmark & \xmark & \cmark & \xmark\\
         \hline
         \hline
         Zhang et al., 2020 \cite{Zhang2020}  & \cmark & \xmark & \xmark & (\cmark) & \cmark\\
         \hline
         Dinh et al., 2022 \cite{Dinh2022}  & \xmark & \cmark & \xmark & \cmark & 
         \xmark \\
         \hline
         Ahmed et al., 2016 \cite{Ahmed2016}  & \cmark & \xmark & \xmark & \xmark & \xmark \\
         \hline
        Gao et al., 2022 \cite{Gao2022}  & \cmark & \cmark & \xmark & \cmark & \xmark \\
         \hline
        López et al., 2019 \cite{López2019}  & \cmark & \cmark & \xmark & \xmark & \xmark \\
        \hline
        \hline
        Present work   & \cmark & \cmark & \cmark & \cmark & \cmark\\
         \hline
    \end{tabular}
    \label{tab_literature}
\end{table*}

Table~\ref{tab_literature} lists the relevant papers from the literature for our study. All the studies combine methods from the field of machine learning with traditional control approaches. Javed et al. \cite{Javed2017}, Kim et al. \cite{Kim2020}, Dey et al. \cite{DEY2023}, Zou et al. \cite{Zou2020} and Dinh et al. \cite{Dinh2022_2} use an intelligent control approach for heating ventilation and air-conditioning (HVAC) systems. Javed et al. \cite{Javed2017} use an ANN to predict the occupancy level and adjust the control system according to the predictions. Kim et al. \cite{Kim2020} train an ANN to model the building behavior, which they then integrate into the optimization problem. The output is a multi-hour schedule. However, no direct control mechanism is based on the ANN, and a forecast is necessary for their approach. Dey et al. \cite{DEY2023} train an RL agent using synthetic data from a rule-based control heuristic as a warm start for the RL training procedure. Zou et al. \cite{Zou2020} use RL with a Long-Short-Term-Memory (LSTM) as the environment for interaction. They train the LSTM to predict the behavior of HVAC systems using historical data from rule-based control. Dinh et al. \cite{Dinh2022_2} solve a mixed-integer linear program (MILP) to generate optimal control actions for an HVAC system. They optimize the electricity costs while maximizing thermal comfort. 

Zhang et al. \cite{Zhang2020}, Dinh et al. \cite{Dinh2022}, Ahmed et al. \cite{Ahmed2016}, Gao et al. \cite{Gao2022} and López et al. \cite{López2019} use machine learning methods to improve the control of different flexible appliances in buildings. Zhang et al. \cite{Zhang2020} use transfer learning for scheduling-based loads and storage systems. Their RL agent learns from a pre-trained agent of another building. Dinh et al. \cite{Dinh2022} control the energy flows of a battery to minimize the costs with a time-variable electricity tariff. They train an ANN with the optimal control actions derived from solving a MILP. Their approach requires a forecast of the future demand, which they compute using a recurrent neural network. Ahmed et al. \cite{Ahmed2016} use an ANN to learn the control actions from a simulation. They apply their approach to household devices that can be switched on and off, like air conditioners, electric water heaters, washing machines, and refrigerators. Gao et al. \cite{Gao2022} train an ANN with the output of a MILP scheduling problem. The ANN generates actions that are then used to solve a simplified MILP problem. This reduces the solving time of the MILP. They consider batteries and diesel generators as flexible devices to minimize energy costs with a time-variable electricity price. López et al. \cite{López2019} use an ANN to control electric vehicle charging to minimize the electricity cost. The labeled data is created by optimally solving the respective control problem using dynamic programming.

\subsection{Contribution}
\label{subsec_Contribution}
To the best of our knowledge, the approach we introduce in this paper is the only one that embeds a forecast-free control approach based on imitation learning from optimal actions into a smart control heuristic. Further, in contrast to all other papers from the literature, our study compares the novel control approach to both optimal control and a positively evaluated smart control heuristic from the literature. Another main feature of our study is using training data from other buildings for the supervised learning approach. In summary, our paper has the following three unique contributions:

\begin{itemize}
\item We introduce a novel imitation learning based control approach that combines domain knowledge with supervised ML. The introduced control approaches are compared to optimal control, smart heuristic control, and conventional control approaches.
\item We test different ML methods for their applicability to improve the control approach of a heat pump. We generate the training data by optimally solving a MILP.
\item Our study investigates the ability of the introduced approach to generalize by applying a model that has been trained with data from other similar buildings.
\end{itemize}

%% file: sections/03_optimization_problem_for_the_building.tex
\section{Optimization problem of the building}
\label{sec:Optimization problem of the building}
To determine the optimal schedule for the heat pump, we solve a MILP. The goal of the building is to minimize the electricity cost $C$ under a time-variable electricity tariff $p_t$. The heat pump utilizes the building mass as thermal storage. Eq.~\ref{eq:objective_function} shows the objective function. The variable $x_t$ determines the modulation degree of the heat pump. Next to the electrical power of the heat pump, the electricity consumption of the inflexible devices $P_t^{DemandEl}$ also influences the objective function.

\begin{equation}
    \min C =  \sum_{t=0}^{Z} (x_t \cdot P^{\stxt{HPmax}} + P_t^{\stxt{DemandEl}}) \cdot \Delta t \cdot p_t \label{eq:objective_function}
\end{equation}
subject to:
\begin{AllowDisplayBreaks}
  \begin{align}
    &T^{\stxt{min}} \leq T^{\stxt{Building}}_t \leq T^{\stxt{max}}
     \quad \forall t 
    \label{eq:temperature_balance} \\
    &T^{\stxt{Building}}_t = T^{\stxt{Building}}_{t-1} + \frac{Q_t^{\stxt{HP}} - Q_t^{\stxt{DemandSH}} -  Q_t^{\stxt{LossesSH}}}{V^{\stxt{UFH}} \cdot  \rho^{\stxt{Concrete}} \cdot c^{\stxt{Concrete}}}  \quad \forall t \neq 0 \label{eq:temperature_difference_equation} \\
    &Q_t^{\stxt{HP}} = x_t \cdot P^{\stxt{HPmax}}  \cdot COP_t \cdot \Delta t \quad \forall t \label{eq:heating_energy_HP} \\
    &h^{\stxt{switchedOff}}_t \leq h^{\stxt{on}}_{t-1} \quad \forall t \neq 0 \label{eq:hp_switched_off_1}  \\
    &h^{\stxt{switchedOff}}_t \leq 1 - h^{\stxt{on}}_{t}   \quad \forall t \label{eq:hp_switched_off_2}  \\
    &h^{\stxt{switchedOff}}_t \geq h^{\stxt{on}}_{t-1} - h^{\stxt{on}}_{t}  \quad \forall t \neq 0 \label{eq:hp_switched_off_3} \\
    &k_t^{\stxt{switchedOff}} =   \sum_{\tau=0}^{t} h_\tau^{\stxt{switchedOff}}   \quad \forall t   \label{eq:hp_switched_off_4} \\
    &k^{\stxt{switchedOff}}_t \leq n^{\stxt{switchedOff}} \quad \forall t\label{eq:hp_switched_off_5} \\
    &x_t \leq h^{\stxt{on}}_{t} \quad \forall t\label{eq:hp_min_mod_1} \\
    &x_t \geq h^{\stxt{on}}_{t} \cdot mod^{\stxt{min}} \quad \forall t \label{eq:hp_min_mod_2} \\
    &x_t \in [0,1], h^{\stxt{switchedOff}}_t \in \{0,1\}, h^{\stxt{on}}_{t} \in \{0,1\} \label{eq:non_negativity} 
  \end{align}
\end{AllowDisplayBreaks}

Eqs.~\ref{eq:temperature_balance} to \ref{eq:non_negativity} define the problem's constraints. Eq.~\ref{eq:temperature_balance} ensures that the temperature of the building $T^{\stxt{Building}}_t$ is always between a lower $T^{\stxt{min}}$ and an upper $T^{\stxt{max}}$ limit. We use a one-zone model for the temperature of the building with the energetic difference equation  Eq.~\ref{eq:temperature_difference_equation}. The heat energy from the heat pump $Q_t^{\stxt{HP}}$ increases the temperature of the building while the demand for space heating $Q_t^{\stxt{DemandSH}}$ and the losses of the heating system itself $Q_t^{\stxt{LossesSH}}$ decrease it. The demand for space heating is a time series incorporating the building's transmission and ventilation losses, as well as internal and solar gains. We describe in Sec.~\ref{subsec:Scenarios_for_the_analysis} how we generated the time series. As we use an underfloor heating system, we need to divide the energetic difference by the volume of the underfloor heating system $V^{\stxt{UFH}}$, the density $\rho^{\stxt{Concrete}}$ and the specific heat capacity of concrete $c^{\stxt{Concrete}}$. To calculate the heating energy of the heat pump, we multiply the heat pump's modulation degree $x_t $ in Eq.~\ref{eq:heating_energy_HP} with the maximum electrical power of the heat pump $P^{HPmax}$, the temperature-dependant coefficient of performance (see \ref{subsec:Scenarios_for_the_analysis}) and the time resolution of the model $\Delta t$. The equation system Eqs.~\ref{eq:hp_switched_off_1} to \ref{eq:hp_switched_off_5} ensures that the number of heat pump starts $k^{\stxt{switchedOff}_t}$ (or switch-offs) does not exceed a predefined upper limit $n^{\stxt{switchedOff}}$. This prevents damage to the compressor resulting from too frequent starts. For this purpose, we use the two auxiliary binary variables $h^{\stxt{switchedOff}_t}$ and $h^{\stxt{on}_{t}}$. The variable $h^{\stxt{switchedOff}_t}$ indicates if the heat pump is being switched off at time $t$ and $h^{\stxt{on}_{t}}$ has the value $1$ if the device is running at time $t$. Eq.~\ref{eq:hp_min_mod_1} and Eq.~\ref{eq:hp_min_mod_2} force the heat pump not to violate a minimum modulation degree $mod^{\stxt{min}}$ while running.

%% file: sections/04_immitation_learning_for_building_control.tex
\section{Imitation learning for building control}
\label{sec:Imitation learning for building control}
 We introduce a novel control approach for heat pumps that combines imitation learning with the intelligent control heuristic \textit{Price-Storage-Control} \cite{FRAHM2023}. Imitation learning is a machine learning paradigm where an agent learns a task by observing and imitating the behavior of an expert demonstrator  \cite{Hussein2017}. In imitation learning, the learning agent aims to replicate the demonstrator's actions to perform a specific task. The agent learns a mapping between observations and actions. In our application, the expert demonstrator consists of actions obtained from optimally solving the optimization problem (MILP) of Sec.~\ref{sec:Optimization problem of the building}. 

\begin{figure}[htb]
    \centering
    \begin{subfigure}[h]{0.48\textwidth}
        \centering
        \includegraphics[height=0.34\textheight]{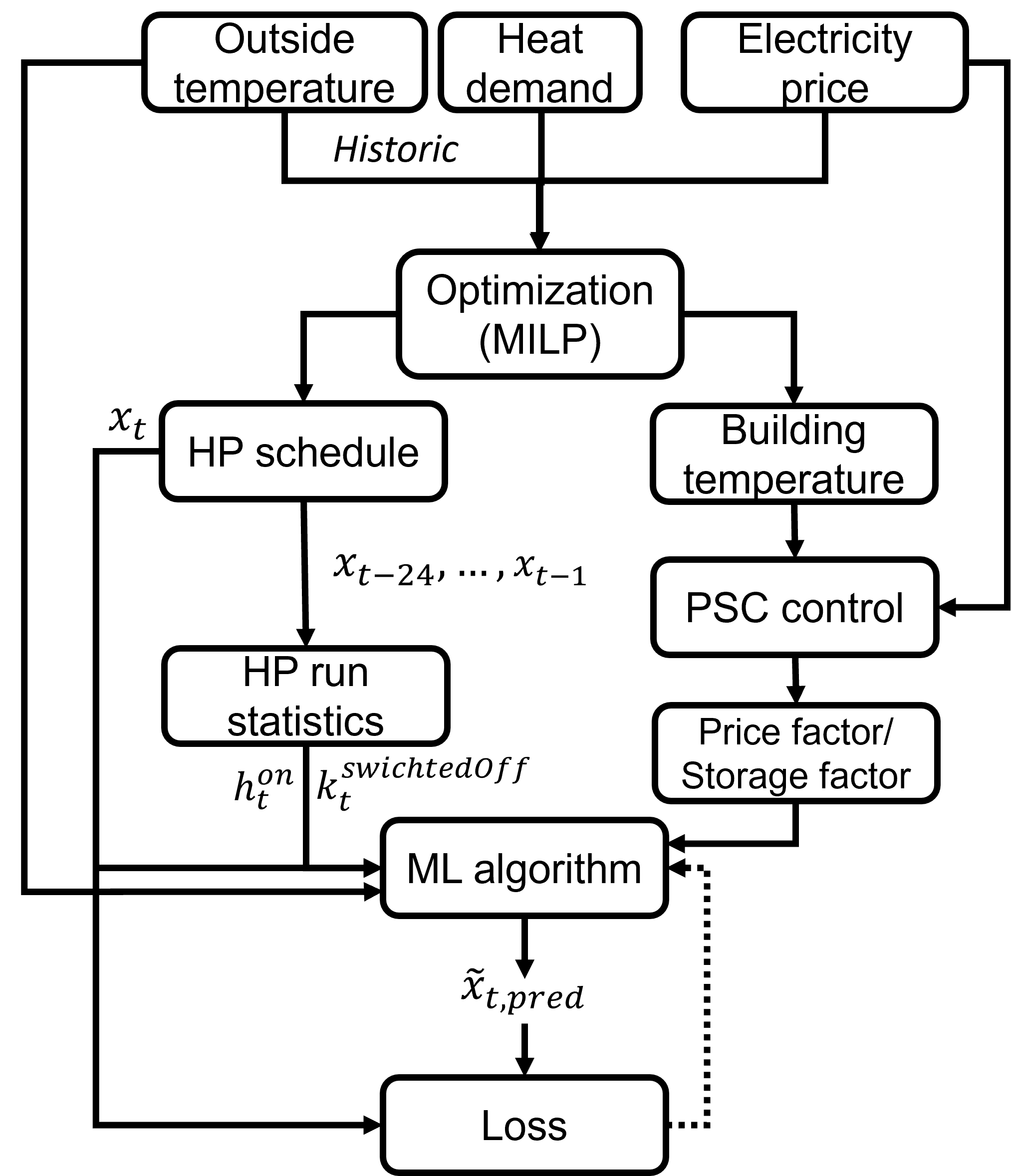} 
    \end{subfigure}
   \caption{PSC-ML training}
    \label{fig:PSC_ML_Training}
\end{figure}

 Fig.~\ref{fig:PSC_ML_Training} illustrates the workflow of our approach to create the heat pump controller \textit{PSC-ML}. We use historical data on electricity prices, heat demand, and outside temperatures as input for an optimization problem. The optimization problem includes a simple building model with a uniform temperature whose difference equation is described in Sec.~\ref{sec:Optimization problem of the building}. The output of the optimization problem is the heat pump's optimal heating actions for every time slot (HP schedule) and the resulting building temperature.

 From the HP schedule of the last 24 hours, two HP run statistics are derived. These statistics indicate how often the heat pump has started the compressor during the current day and if the heat pump was running at time $t$. This information serves as one part of the input features for the machine learning (ML) algorithm.

 The other part of the input features are the \textit{Price factor} and the \textit{Storage factor} calculated by \textit{PSC} algorithm using the resulting building temperature from the optimization problem and the historic electricity price. The \textit{Price factor} quantifies for every time slot the share of future electricity price values in the next 24 hours that are higher than the current price $p_t$. The \textit{Storage factor} determines how far the current temperature of the building $T^{Building}_t$ is away from the lower $T^{min}$ and upper $T^{max}$ temperature limits. Details about the calculations of the control algorithm \textit{PSC} are explained in \cite{FRAHM2023}. Additionally, we use the outside temperature as another input feature.

The optimal heating action $x_t$ from the resulting HP schedule serves as the target label for the ML algorithm, which we obtain by solving the optimization problem. The ML algorithm is trained to capture the relationship between the inputs of the control problem and the optimal output. It essentially derives a regression model that maps the inputs to the optimal outputs. We calculate the loss during the training by comparing the predicted output of the ML algorithm $ \widetilde{x}_{t,pred} $ with the corresponding optimal output from the optimization problem $x_t$. The trained model is subsequently incorporated into the heat pump and integrated with the \textit{PSC} algorithm as part of the novel control approach \textit{PSC-ML}. 

\begin{figure}[htb]
    \centering
    \begin{subfigure}[h]{0.48\textwidth}
        \centering
        \includegraphics[width=\textwidth]{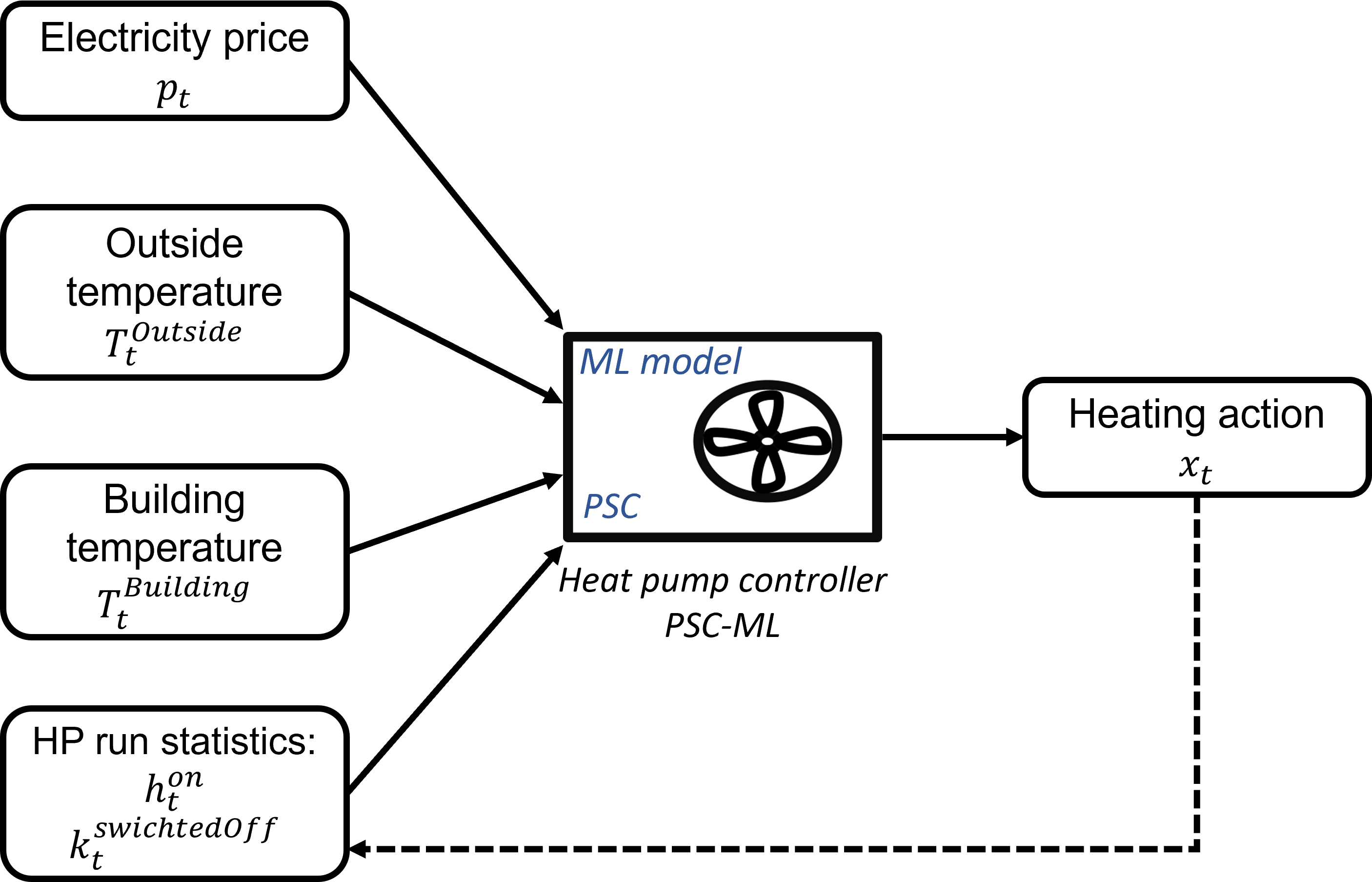}

    \end{subfigure}
   \caption{PSC-ML applied}
    \label{fig:PSC_ML_Applied}
\end{figure}

 Fig.~\ref{fig:PSC_ML_Applied} illustrates the application of \textit{PSC-ML}. At every time slot $t$, the heat pump controller observes the relevant input data. The \textit{PSC} algorithm and the trained ML model are integrated into the heat pump. The HP controller uses the outside temperature $T^{Outside}_t$, the number of starts of the heat pump so far $ k_t^{\stxt{switchedOff}}$, and the binary variable $h^{\stxt{on}}_{t}$ directly. Using the electricity price $p_t$ and the building temperature $T^{Building}_t$, the \textit{PSC} algorithm derives the \textit{Price factor} and the \textit{Storage factor}. These five features are the input for the trained ML regression model, which outputs the heating action $x_t$, influencing the HP run statistics during the next time slot. The \textit{PSC-ML} controller includes some super-ordinate rules that can overrule the actions proposed by the ML algorithm. These rules are similar to the ones explained in \cite{DENGIZ2019}. They ensure that the heat pump heats the building if the room temperature is too low and stops heating if it is too high. 

\begin{figure}[htb]
    \centering
    \begin{subfigure}[h]{0.48\textwidth}
        \centering
        \includegraphics[width=\textwidth]{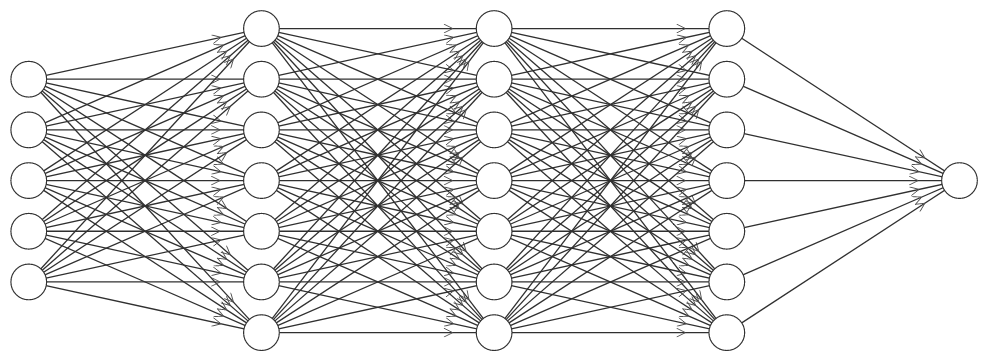}

    \end{subfigure}
   \caption{ANN schema}
    \label{fig:ANN_Schema}
\end{figure}

We use different machine learning approaches (see Section \ref{sec:Results}), but applying an ANN as the machine learning approach yields the best results. Fig.~\ref{fig:ANN_Schema} illustrates a schematic view of the ANN. The ANN is a multi-layer perceptron with fully connected nodes and the rectified linear unit (ReLU) as the activation function. As described in Section \ref{sec:Results}, we investigated different configurations and hyperparameters.

We use three types of buildings that differ regarding their insulation level and, thus, their heat demand. Heat pumps are mainly used in buildings with good insulation standards due to increased efficiency. Hence, the analysis includes buildings with a heat demand of 25 $kWh$, 50 $kWh$, and 80 $kWh$ per year for every square meter of living area. Overall, the buildings have a good insulation level compared to the average heat demand of buildings in Germany, which lies between $125 \frac{\text{kWh}}{\text{m}^2}$ and $150 \frac{\text{kWh}}{\text{m}^2}$\cite{verbraucherzentrale2023} per year. However, these values are typical for new buildings in Germany.

\begin{figure}[htb]
    \centering
    \begin{subfigure}[h]{0.48\textwidth}
        \centering
        \includegraphics[width=\textwidth]{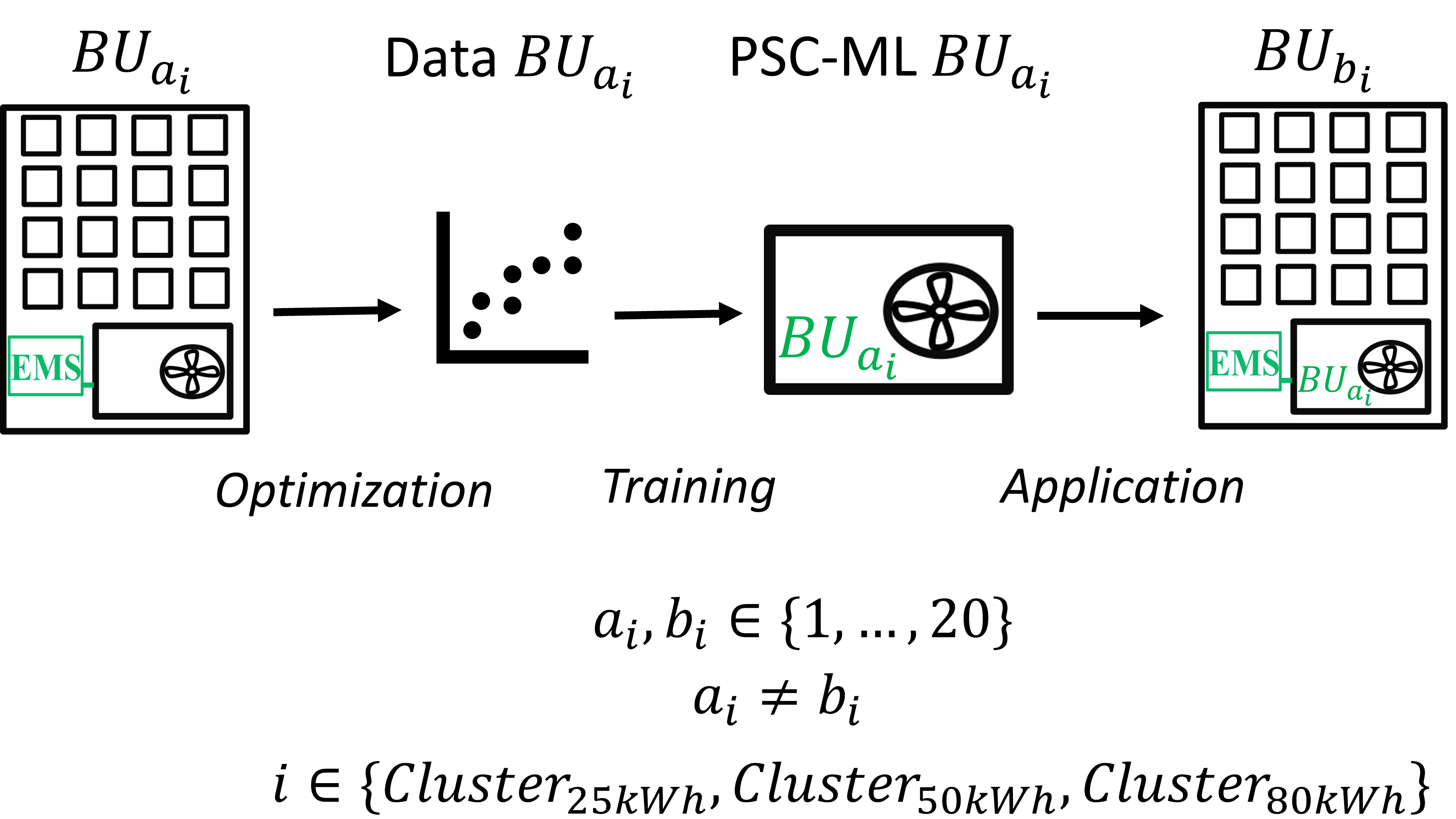}

    \end{subfigure}
   \caption{Training process with data from other buildings}
    \label{fig:Training_with_other_buildings}
\end{figure}

Our analysis includes data from different buildings for training the ANN to see if a trained model can be applied to similar buildings. Fig.~\ref{fig:Training_with_other_buildings} illustrates the training process. We use building $BU_{a_i}$ from cluster $i$ of buildings to generate the training data by solving the optimization problem for that building. Each cluster includes buildings with a similar heat demand per year and square meters of living area. For each building cluster, we created 20 different datasets (see Section~\ref{subsec:Scenarios_for_the_analysis}). The time series of heat demand within each cluster varies regarding the temporal dimension and the magnitude of values. These differences arise due to temporal shifts and multiplication of the time series with a constant factor, leading to variations in both the timing and the magnitude of the heat demand data. The resulting average yearly heat demand for every building within one cluster differs at most $+5 \frac{\text{kWh}}{\text{m}^2}$ or $-5 \frac{\text{kWh}}{\text{m}^2}$ from the base demand, which is $25 \frac{\text{kWh}}{\text{m}^2}$, $50 \frac{\text{kWh}}{\text{m}^2}$ and $80 \frac{\text{kWh}}{\text{m}^2}$ for the corresponding clusters. 

The generated training data for building $BU_{a_i}$ is then used as input for the ANN. This results in the specific controller \textit{PSC-ML} for building $BU_{a_i}$. To test the capability of our approach to generalize and thus to be applied in different buildings, we use another building from the same cluster $BU_{b_i}$ for applying the novel control approach. Thus, building $BU_{b_i}$ uses the controller \textit{PSC-ML $BU_{a_i}$} for the evaluation that has been trained with data from another building.

%% file: sections/05_results.tex
\section{Results}
\label{sec:Results}

\subsection{Scenarios for the analysis}
\label{subsec:Scenarios_for_the_analysis}

We use the air-source heat pump \textit{Compress 6800i AW} \cite{bosch_compress6800} from the company \textit{Bosch Thermotechnik} with a maximum electrical power of $3000~W$ and a minimal modulation degree of $20\%$. For the efficiency factor $COP$, we assume a linear relationship between the values provided in the technical documentation of the heat pump model as in \cite{DENGIZ2019}. The efficiency $COP_t$ for a time slot $t$ depends on the difference between the sink temperature, which is the underfloor heating system's supply temperature of $30~^\circ C$, and the source temperature (outside temperature). For the $25~kWh$ buildings, we aggregated three of the mentioned heat pump models into one with an electrical power of $9000~W$, while the $50~kWh$ buildings have four heat pumps combined in cascading operations, and the $80~kWh$ have six. We set the maximum number of starts for the heat pump to 28 during one week. 

The buildings are all multi-family houses located in the federal state of \textit{Schleswig-Holstein} in the north of Germany. The building has 12 apartments, each with a living area of $75~m^2$ and a concrete width of $7~cm$ for the UFH \cite{DENGIZ2019}. The lower limit for the temperature is $20.5~^\circ C$, while the upper limit is at $23.5~^\circ C$. The losses of the UFH were assumed to be $45~W$. It has to be noted that these are merely the losses of the heating system's tubes that do not contribute to heating the building. The heat demand data incorporates the building's significantly higher transmission heat losses.

 We use the CREST model \cite{RICHARDSON20101878} and German time use data \cite{destatis_2024} for occupancy behavior simulation, which builds the basis for the consistent simulation of building space heating and electricity demand. Appliance starts are simulated based on the occupancy profiles, local weather conditions, and household device equipment. Finally, the appliance-specific load profiles are aggregated to one electricity demand profile per household. Additionally, the 5R1C thermal building model \cite{ISO13790} is used to calculate the demand for space heating based on occupancy-related internal heat gains and thermal building parameters taken from the TABULA residential building typology \cite{IWU_2015}.

For the electricity price, we use the data of the Day-Ahead market in Germany from 2021, which we took from the \textit{ENTSO-E Transparency Plattform} \cite{entsoe_2023}. We scaled them up to align with the average electricity price for residential customers in Germany. The prices have an hourly resolution.

\subsection{Evaluation}
\label{subsec_Evaluation}

We use the optimal schedules of 20 weeks to train the ML models. The time resolution of the data obtained from the optimization is 30 minutes. We split the training data into $70\%$ for training the model and  $30\%$ for the validation. After training, the ML model, embedded into the control approach \textit{PSC}, is tested in 20 weeks that were not included in the training data. For every week in the evaluation, we train the model from scratch and choose 20 weeks of training data randomly from 26 weeks of available training data (we only considered the heating period from October to March). The time resolution of the control actions $\Delta t$ (and thus of the training data) is 30 minutes.

For our first evaluation, we train the model with data from one building and apply the trained model to 20 different buildings from the same building cluster. The average training time per week was 9 minutes using the GPU \textit{Nvidia GeForce RTX 3090} \cite{nvidia_2024} and an \textit{Intel Core i9-7940X} \cite{intel_2024} as CPU. The average time for deriving and executing the control actions for one week (2 decisions per hour) is 3 seconds with the trained \textit{PSC-ML} control approach, while the exact optimization needs about 40 seconds. We use the Gurobi solver \cite{gurobi} to solve the optimization problem of the building.

\begin{figure}[htb]
    \centering
    \begin{subfigure}[h]{0.49\textwidth}
        \centering
        \includegraphics[width=\textwidth]{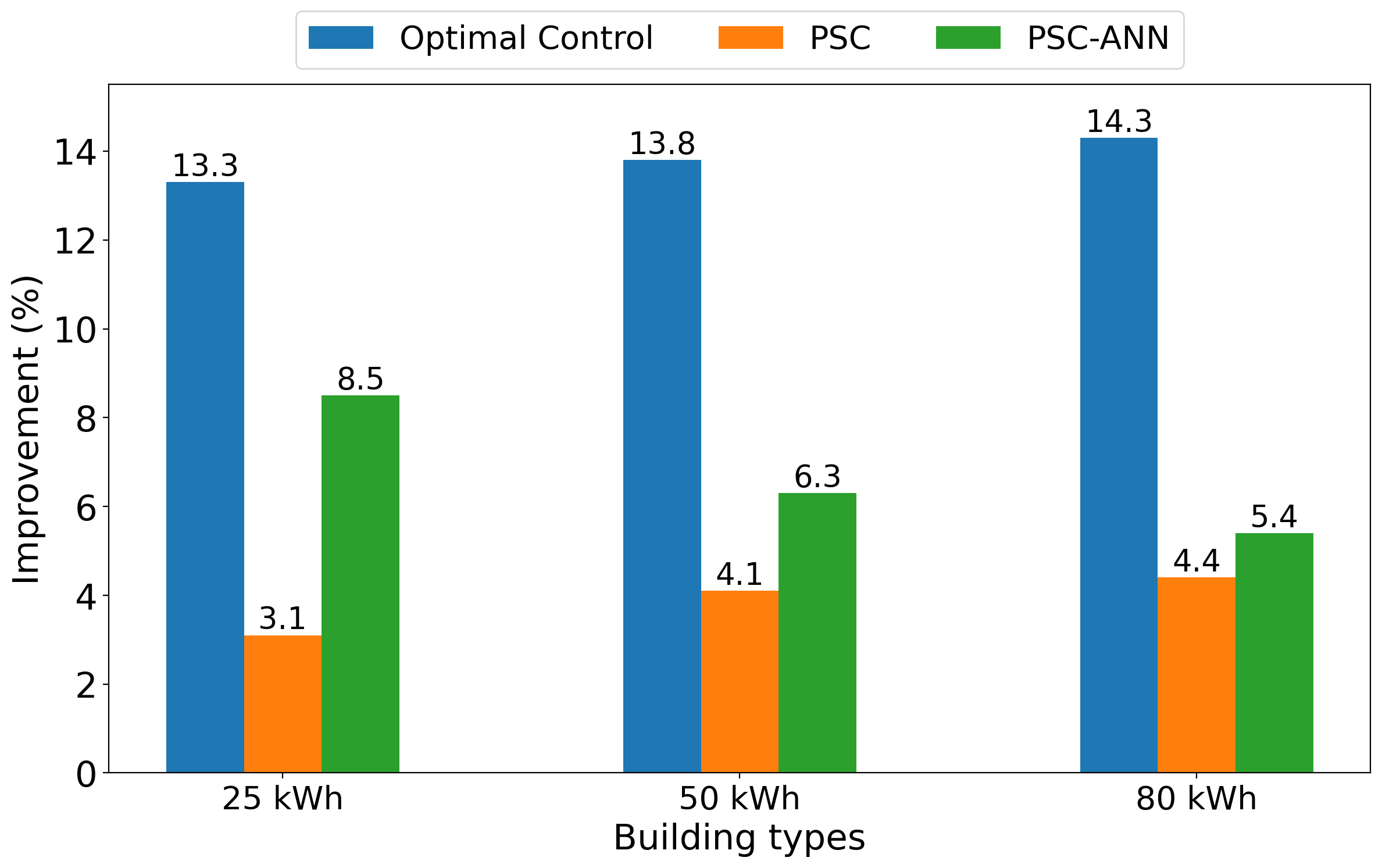}

    \end{subfigure}
   \caption{Improvement regarding the cost reduction compared to the Conventional Control approach averaged over 20 buildings per building type for 20 weeks and three runs}
    \label{fig_Result_Percentage_Improvements}
\end{figure}

Fig.~\ref{fig_Result_Percentage_Improvements} depicts the cost reduction of the methods \textit{Optimal Control}, \textit{PSC} and \textit{PSC-ANN} compared to \textit{Conventional Control}. We obtained the results for the three building type clusters by averaging over 20 buildings per building type and 20 weeks. We further made three simulation runs per method. Thus, each bar represents the average values of 1200 evaluation weeks. The results reveal that for all building-type clusters, the three methods lead to cost improvements compared to the Conventional Control approach. As expected, the \textit{Optimal Control} yields the best results because it is based on the assumption of having a perfect forecast of future demands and prices (or, equivalently, calculating the optimal control actions retrospectively using historical data). Thus, the \textit{Optimal Control} merely represents a theoretical upper bound for the improvements. In contrast, \textit{PSC} and \textit{PSC-ANN} do not require a building model or any forecast to derive their control actions. The diagram shows that \textit{PSC-ANN} outperforms \textit{PSC} for all building types. The additional cost reduction using an ANN to calculate the control actions ranges between $1\%$ for the 80 kWh buildings and $4.7\%$ for the 25 kWh buildings.

Table~\ref{table_averages_costs} lists the average electricity costs per week for the 50 kWh buildings averaged over 20 buildings and three runs. \textit{PSC} and \textit{PSC-ANN} lead to significantly better results than \textit{Conventional Control}. While in three out of 20 weeks, \textit{PSC} leads to lower costs compared to \textit{PSC-ANN}, on average, the use of \textit{PSC-ANN} leads to cost reductions of more than $5~$€ per week. 

\begin{table}[htbp]
  \centering
  \caption{Average electricity costs in € per week for the 50 kWh buildings averaged over 20 buildings and three runs}
  \begin{tabular}{lcccc}
    \toprule
    Week & \makecell{Optimal \\ Control} & \makecell{Conventional \\ Control} & PSC & PSC-ANN \\
    \midrule
    1 & 269.52 & 323.33 & 316.00 & 317.91 \\
    2 & 155.79 & 205.12 & 189.73 & 178.63 \\
    3 & 273.40 & 316.44 & 306.49 & 295.91 \\
    4 & 178.59 & 200.89 & 198.92 & 185.42 \\
    5 & 325.30 & 376.47 & 356.56 & 356.41 \\
    6 & 228.01 & 270.14 & 256.62 & 246.17 \\
    7 & 281.37 & 324.84 & 306.51 & 300.89 \\
    8 & 233.47 & 272.67 & 257.95 & 257.00 \\
    9 & 133.24 & 154.66 & 153.04 & 147.75 \\
    10 & 186.08 & 217.42 & 205.24 & 201.86 \\
    11 & 178.45 & 201.28 & 192.80 & 185.83 \\
    12 & 188.51 & 214.73 & 206.55 & 204.79 \\
    39 & 122.47 & 129.85 & 128.52 & 125.07 \\
    40 & 135.51 & 146.05 & 146.30 & 136.90 \\
    41 & 140.41 & 158.68 & 157.07 & 145.00 \\
    42 & 254.95 & 292.63 & 276.92 & 266.73 \\
    43 & 146.68 & 155.92 & 155.54 & 152.18 \\
    44 & 241.43 & 276.34 & 263.73 & 264.05 \\
    45 & 221.62 & 272.17 & 253.56 & 257.25 \\
    46 & 153.90 & 188.46 & 177.96 & 175.29 \\
    \midrule
    Average & 202.44 & 234.90 & 225.30 & 220.05 \\
    \bottomrule
  \end{tabular}
  \label{table_averages_costs}
\end{table}

\begin{figure}[htb]
    \centering
    \begin{subfigure}[h]{0.49\textwidth}
        \centering
        \includegraphics[width=\textwidth]{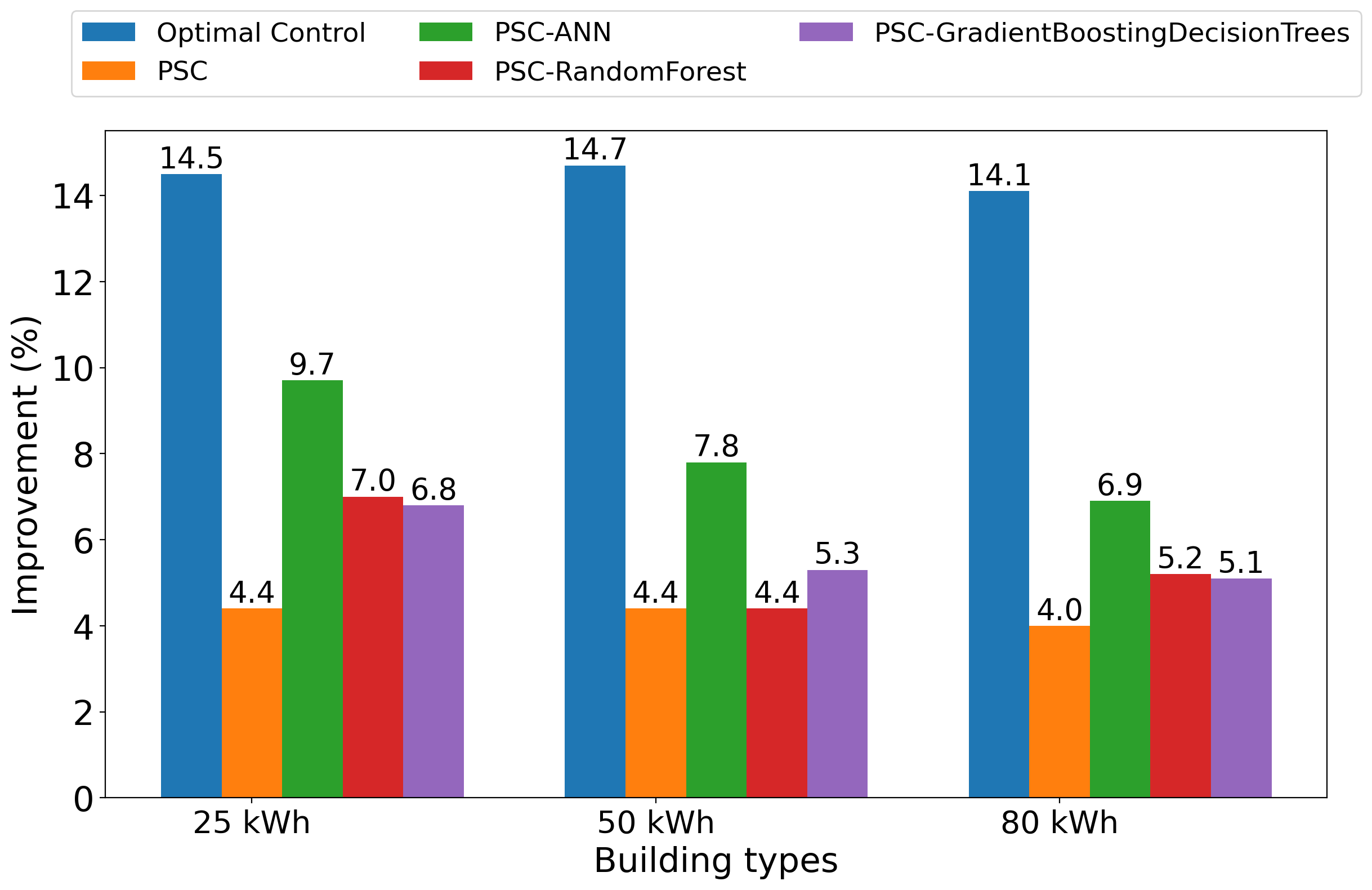}

    \end{subfigure}
   \caption{Average improvement compared to the Conventional Control approach for different machine learning methods averaged over 6 buildings per building type for 20 weeks}
    \label{fig_Result_Improvement_Multiple_Methods}
\end{figure}

 Next to an artificial neural network, we investigate the application of Random Forest and Gradient Boosting decision trees as the machine learning approach. Fig.~\ref{fig_Result_Improvement_Multiple_Methods} shows the results of the experiments. The bars quantify the average improvement compared to the \textit{Conventional Control} approach for different machine learning methods averaged over six buildings per building type for 20 weeks. The diagram reveals that applying any function approximator improves the results of \textit{PSC}. Using artificial neural networks for imitation learning for all three building types leads to the most significant improvements and, thus, the lowest costs in our experiments. Because of this, we only use ANN for our large-scale analysis with 20 different buildings per building type (see Fig.~\ref{fig_Result_Percentage_Improvements}).

\begin{figure}[htb]
    \centering
    \begin{subfigure}[h]{0.49\textwidth}
        \centering
        \includegraphics[width=\textwidth]{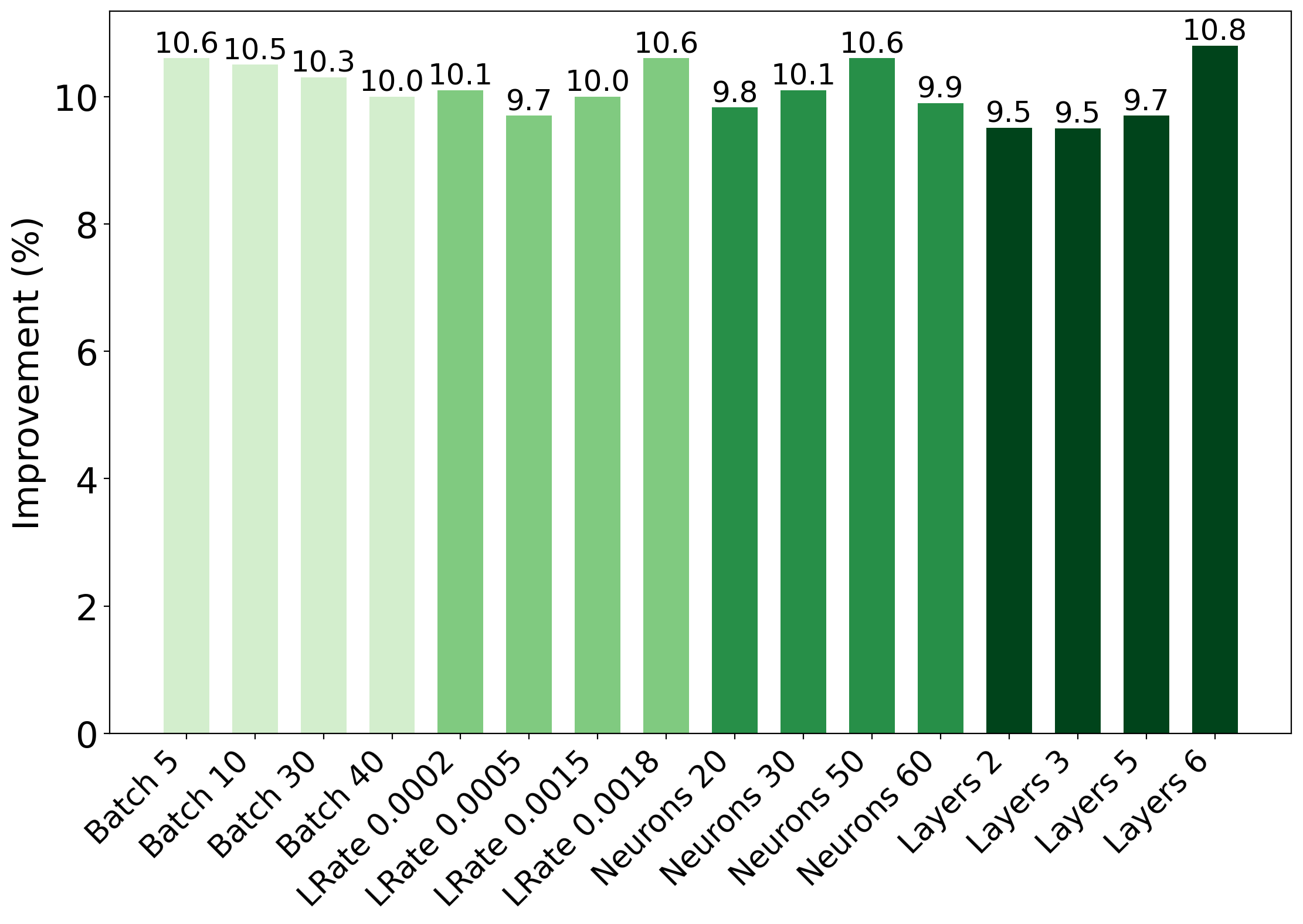}
        
    \end{subfigure}
   \caption{Improvement compared to the Conventional Control approach for different ANN parameters averaged over one 25 kWh building for 20 weeks}
    \label{fig_Result_Improvement_ANN_hyperparameters}
\end{figure}

We ran experiments with different hyperparameters of the ANN. Fig.~\ref{fig_Result_Improvement_ANN_hyperparameters} illustrates the results of our analysis for one 25 kWh building averaged over 20 weeks. We varied the batch size, the learning rate, the number of neurons per layer, and the number of layers. In general, the impact of the different parameters is mediocre, ranging from improvements of $9.5\%$ to $10.8\%$. For our large-scale experiment, we chose the hyperparameter combination that yielded the best results (batch size 30, learning rate 0.0018, number of neurons per layer 50, number of layers 5).

\begin{figure*}[htb]
    \centering
    \begin{subfigure}[h]{0.99\textwidth}
        \centering
        \includegraphics[width=\textwidth]{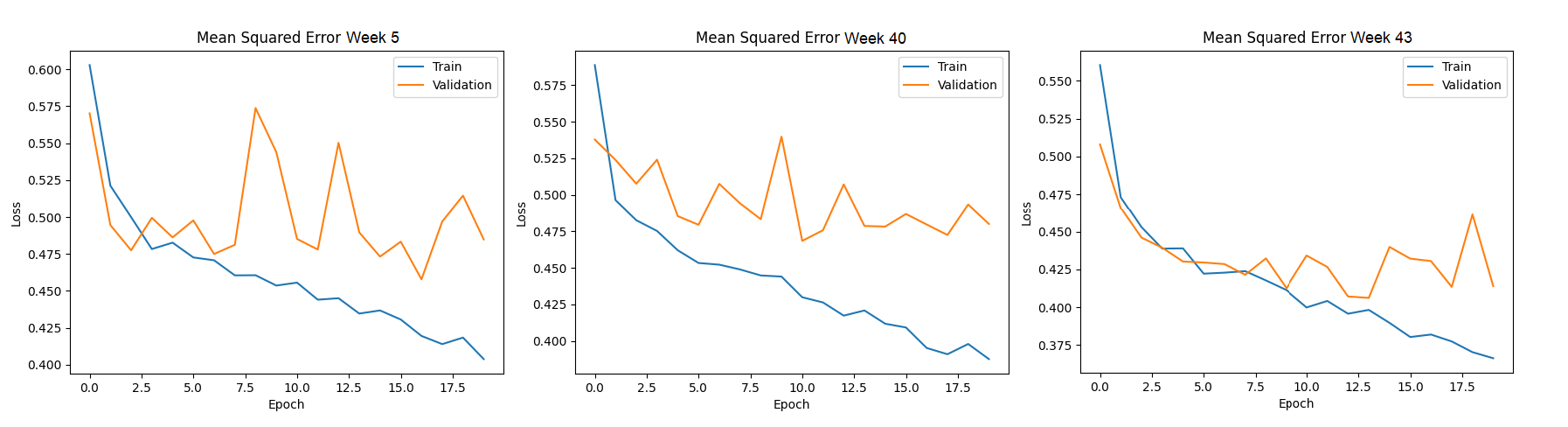}
        
    \end{subfigure}
   \caption{RMSE as a function of epochs for three weeks of a 25kWh building}
    \label{fig_3_weeks_rmse}
\end{figure*}

Fig.~\ref{fig_3_weeks_rmse} plots the loss function (mean squared error) as a function of the epochs of the training and validation dataset for three weeks of a 25kWh building. We choose 20 epochs as the error function in the validation dataset does not significantly reduce after a certain number of epochs. We observed this in almost all weeks and for multiple different buildings. 

We implemented the simulations and the optimization in the programming language \textit{Python}. For the MILP, we use the package \textit{Pyomo} \cite{hart2011pyomo}, for the ML approaches \textit{keras} \cite{chollet2015keras} and \textit{scikit-learn} \cite{scikit-learn}.

\subsection{Critical appraisal}
\label{subsec:Critical appraisal}
To carry out our experiments, we made some simplifications. The used building model is relatively simple as it only has one temperature zone. Thus, it does not consider internal heat flows and solar heat gains. The solar heat gains are incorporated in the demand data. We also did not analyze the use of domestic hot water tanks as an additional source of flexibility needed throughout the year. Further, we solely considered one building with one flexible device. The analysis of coordinated reactions to demand response signals from multiple buildings was not in the scope of this paper. We carried out all our investigations in a simulation environment. Thus, we did not apply our proposed approach in real-world experiments. Our method defines a general procedure to combine domain knowledge with machine learning for control problems of electrical heating devices. Because of this, we are convinced that our developed control algorithm can be used for more complex building settings and real-world applications.

%% file: sections/06_conclusion.tex
\section{Conclusion}
\label{sec:Conclusion}

In this paper, we developed a novel control approach for demand response with heat pumps \textit{PSC-ANN} based on imitation learning. It uses the effective control method \textit{PSC} from the literature and trains an artificial neural network with optimal control action data. The trained model quickly derives the control actions based on real-time data. The results reveal the benefits of including an artificial neural network in the control process. The introduced approach outperforms an intelligent control algorithm that has been positively evaluated. Further, we showed that the trained model can generalize and thus be used for various similar buildings without the need for training again with building-specific data. 

This enables our developed approach to be applied in many buildings and scenarios. All tested models strongly outperform a conventional control approach. They lead to a better utilization of volatile renewable energy sources and can also be used to stabilize the electricity grid. 

Future work will consider multiple buildings with intelligent controllers and their interactions. Methodologically, the use of machine learning methods for dealing with sequential data, like recurrent neural networks, constitutes a promising approach for building-related control problems. In addition, we plan to compare our approach to methods from the field of reinforcement learning, especially regarding the sample efficiency. Also, we intend to include other flexibility options like electric vehicles, hot water tanks, or battery storage systems in our analysis.

%% file: sections/ending.tex
\section*{Supplementary materials}
\label{sec_supplementary_materials}
Both the code and data are openly accessible. The repository containing the commented code is available at 
\href{https://github.com/thomasdengiz/Imitation_Learning_Weeks}{GitHub}. 
The input data and result profiles can be accessed at 
\href{https://publikationen.bibliothek.kit.edu/1000172089}{KITOpen}.

\section*{Appendix}
\label{sec_appendix}

\begin{figure*}[t]
    \centering
    \begin{subfigure}[h]{1.0\textwidth}
        \centering
        \includegraphics[width=\textwidth]{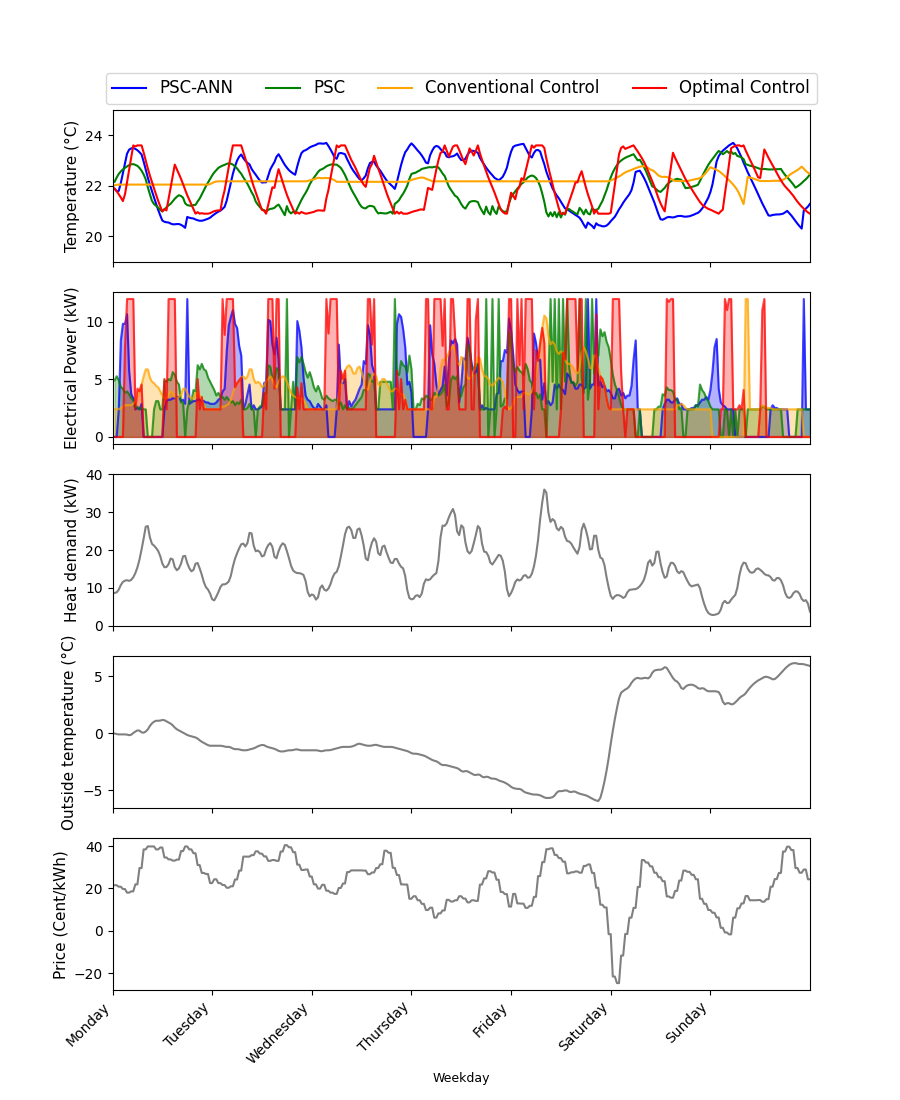}

    \end{subfigure}
   \caption{Exemplary profiles for one week in December for a 50 kWh building}
    \label{fig_Resulting_Profiles_One_Week}
\end{figure*}

Exemplary profiles for one week resulting from the different control approaches are presented in Fig.~\ref{fig_Resulting_Profiles_One_Week}.

\section*{CRediT author statement}
\label{sec_Credit_author_statement}
\textbf{Thomas Dengiz}: Conceptualization, Methodology, Software, Validation, Formal analysis, Investigation, Writing - Original Draft, Visualization. \textbf{Max Kleinebrahm}: Validation, Resources, Data Curation, Writing - Original Draft

\section*{Acknowledgments}
\label{sec_acknowledgments}
This work was supported by the project AsimutE (Autoconsommation et Stockage Intelligents pour une Meilleure Utilisation de l’Énergie) from the European Territorial Cooperation program Interreg.